\documentclass[11pt,showpacs,preprintnumbers,amsmath,aps,amssymb]{revtex4}
\usepackage{graphicx}
\usepackage{dcolumn}
\usepackage[english]{babel}
\usepackage{bm}
\usepackage[font=scriptsize]{caption}
\usepackage{ragged2e}
\usepackage[font=small,labelfont=bf,justification=raggedright]{caption}
\usepackage[utf8]{inputenc}
\usepackage[tikz]{bclogo}
\usepackage[framemethod=tikz]{mdframed}
\usepackage{amsmath}

\usepackage{mathtools}
\usepackage{xcolor}
\usepackage{lscape}
\usepackage{cancel}
\usepackage{rotating}
\colorlet{BLUE}{blue} \colorlet{RED}{red}
\newsavebox{\measurebox}
\usepackage{subcaption}

\begin{document}
\title{F(R,..) theories from the point of view of the Hamiltonian approach: non-vacuum Anisotropic Bianchi type I cosmological model.}

\author{J. Socorro$^1$ }
\email{socorro@fisica.ugto.mx}
\author{Juan Luis P\'erez $^1$ }
\email{jl.perezp@ugto.mx}
\author{Luis Rey D\'iaz-Barr\'on$^2$}
\email{rdiaz@ipn.mx}
\author{Abraham Espinoza Garc\'ia$^2$}
\email{aespinozag@ipn.mx}
\author{Sinuh\'e P\'erez Pay\'an$^2$}
\email{aperezp@ipn.mx}

\affiliation{ $^1$ Department of Physics,  Division of Science and
Engineering, University of Guanajuato, Campus Le\'{o}n,
     Le\'{o}n 37150,  M\'exico,\\
    $^2$ Unidad Profesional Interdisciplinaria de Ingenier\'ia, Campus Guanajuato del Instituto Polit\'ecnico Nacional,
Av. Mineral de Valenciana No. 200, Col. Fraccionamiento Industrial
Puerto Interior, Silao de la Victoria 36275, Guanajuato, Mexico}
\begin{abstract}
In this study, we investigate the implications of F(R) gravity within a classical framework, utilizing an anisotropic Bianchi type I cosmological model. The analysis incorporates standard matter represented by a barotropic fluid with n equation of state $P=\gamma\rho$. We derive classical solutions under two distinct gauges, $N=1$ and $N=6\eta^3 D$, yielding results that are frecuently employed as ansatze for solving the Einstein field equations. For a comprehensive analysis, vacuum solutions are also provided. Furthermore, based on the temporal evolution of the cosmic volume and auxiliary function $\rm D=\partial F/\partial R$, we propose that inflation emerges from geometric properties rather than a fundamental scalar field $\phi$. In this view, the auxiliary function itself facilitates the inflationary phase of the universe's evolution.\\

\end{abstract}

\maketitle
\section{introduction}
Currently the is a prevailing perspective that a modified gravity theory --representing a classical generalization of general relativity-- should be able of consistently describing both early-time inflation and late-time acceleration without the necessity of additional dark coponents \cite{phys-reports-2011}. Such frameworks may provide resolutions to several open questions in modern cosmology, including the nature of dark matter and dark energy, the coincidence problem, the transitions from cosmic deceleration to acceleration, the unification of fundamental interactions and the hierarchy problem. The landscape of modified gravity is extensive, encompassing: $F(R)$
theories \cite{2,6,IJGMMP-2007,7,9,Shamir,3,21,8,1,44,11,12,5,4,14}, $F({\cal G})$ theories \cite{PLB-2005,JPA-2006},  $F(R,{\cal G})$ models \cite{PRD-2006}, $F(R,T)$ and modifications
\cite{grg45-2013,ass357-2015,epjc76-2016,epjc76-449,ass361-2016,prd97-2018,epjc135-2020,universe272-2024}, string-inspired models \cite{PRD-2005,leonel1,leonel2,leonel3} coupled with scalar-Einstein-Gauss-Bonnet gravity \cite{PLB2-2005,CQG-2005}, non-local gravity \cite{63,65} with Gauss-Bonnet term \cite{67,68},
gravity non-minimally coupled with the matter lagrangian
\cite{plb599-2004,prl92-2004,prd72-2005,prd75-2007,epjc70-2010,cqg29-2012,ijtp52-2013,epjc75-2015,epjc81-2021,plb831-2022,plb866-2025},
non-minimally coupled vector model \cite{75}, non-minimal Yang-Mills
theory \cite{77,78}, modified $F(R)$ Horava-Lifshitz gravity
\cite{79}, dark fluid with an inhomogeneous equation of state
\cite{33,34,35,36,119}, covariant power counting renormalizable
gravity \cite{114}, and extensions of these theories
\cite{20,22,27,23,24,25,26,28}. Despite this variety, the complete gravitational action remains an open problem in high energy physics, as it should ideally be derived from a fundamental theory. In the absence of a definitive quantum gravity framework, modified gravity serves as a phenomenological approach, constructed and refined through validation against experimental data or indirect observations.

$F(R)$ theories generalize the Einstein-Hilbert action by replacing the Ricci scalar $R$ with an arbitrary function $F(R)$ \cite{Farasat}. Their utility lies in the versatility, potentially accounting for the universe's expansion at various stages without invoking dark sectors \cite{Rishi}.

While the $\Lambda$CDM model typically assumes a homogeneous and isotropic universe, high resolution Cosmic Microwave Background (CMB) measurements \cite{Mississippi} suggest asymmetries in large-scale expansion \cite{Tripathy}. Consequently, Bianchi Type I models—which describe a spatially flat but anisotropic universe with direction-dependent expansion rates—provide a robust framework for investigating these observed anisotropies within $F(R)$ gravity. In these models, asymmetric expansion is characterized by distinct scale factors ($A,B,C$) within the line element.

Motivated by previous studies of Bianchi Type I vacuum models \cite{Sharif2009, Sharif2010}, this work departs from standard power-law assumptions (e.g., $F\propto a^m)$ found in literature \cite{Farasat, Hasmani}. Instead, we employ the Hamiltonian formalism to analyze our model. We also address the ambiguity regarding the introduction of matter in modified gravity. While some approaches incorporate the matter Lagrangian density directly within the F(...) function \cite{prd75-2007}, others treat it as an independent component \cite{ijtp52-2013} or utilize a hybrid approach \cite{epjc81-2021}. In this work, we present a generalized theory $F(R,T, {\cal
L}_{matter}) \to  F(R,T)\to F(R)$; this includes a general Lagrangian density dependent on metric elements, multi-chiral scalar fields, and K-essence fields. As a "toy model" to demonstrate the efficacy of Hamilton’s equations without relying on an ansatz, we solve the anisotropic equations for the $F(R)$ case without scalar fields; the inclusion of scalar fields and more complex $F$ functions is reserved for future work.

The structure of our work is as follows: Section~\ref{section2} is devoted to derive the equations of motion via variation of the Lagrangian density and analyze the Bergmann-Wagoner theories as a benchmark \cite{Clifford}. In Section~\ref{section3} an analysis of the F(R) toy model is presented, showing the exact solutions for the Bianchi Type I model in both vacuum and perfect fluid contexts using the Hamiltonian formalism. While in Section~\ref{section4}, we determine the Hamiltonian density and the field equations for the gauges N=1 and $N=6\eta^3 D$. We demonstrate that when scale factors and the auxiliary $D$ function are introduced into the field equations, the resulting constraints yield a vanishing Ricci scalar ($R=0$) and, consequently, $F(R)=0$. Finally, in Section~\ref{section5} provides concluding remarks.

\section{Generalized F(R,T,${\cal L}_{matter}$) theory with multiscalar fields and matter content.}\label{section2}
We start considering a gravitational action incorporating a chiral scalar field, a cosmological term, and an ordinary matter content within the framework of $F(R,T,{\cal L}_{matter})$ theory. The action is defined as
\begin{equation}
\rm S=\int \sqrt{-g}\left[ A(\phi_j)F(R,T,{\cal L}_{matter})- 2
\Lambda - M^{ab}(\phi_j)G(\xi_{ab})+C(\phi_j) + {\cal
L}_{matter}\right]d^4x, \label{accion}
\end{equation}
where R is  the Ricci scalar, T is the trace of the energy momentum
tensor, $A(\phi_j)$ is a function of the scalar field, $\Lambda$ the
cosmological constant, $\rm C(\phi_j)$ has the information of a
scalar field potential, $M^{ab}(\phi_j)$ is a matrix for multifield
theory, $G(\xi_{ab})$ depend of the kinetic energy (like K-essence
term) and
\begin{equation}
\rm \xi_{ab}(\phi_j,g^{\mu \nu})=-\frac{1}{2}g^{\mu \nu}\nabla_\mu \phi_a \nabla_\nu \phi_b,
\end{equation}
in addition, the variation of the Lagrangian density ${\cal L}_{matter}$ with respect to the metric yields the energy-momentum tensor 
\begin{equation}
T_{\mu \nu}= -\frac{2}{\sqrt{-g}}\frac{\delta \sqrt{-g}{\cal
L}_{matter}}{\delta g^{\mu \nu}}. \label{tensor}
\end{equation}

For a perfect fluid, the energy-momentum tensor and the corresponding conservation law are
\begin{equation}
T_{\mu \nu} = (\rho + p) u_\mu u_\nu + g_{\mu \nu} p, \qquad \nabla_\nu T^{\mu \nu} = 0, \label{density}
\end{equation}
where $\rho$ is the energy density, P is the fluid pressure in the comoving frame, and $u_\mu$ is the four-velocity. For a barotropic fluid satisfying the equation of state $P=\gamma \rho$ (where $\gamma$ characterizes different cosmological epochs), the Lagrangian density is expressed as
\begin{equation}
\mathcal{L}_{matter} = 16\pi G_N \rho, \label{lagra-matter}
\end{equation}
where in Brans-Dickie-type scalar-tensor theories, the Newton gravitational constant $G_N$ is associated with the scalar field via the relation $G_N \to \phi^{-1}$. Notably, as demonstrated in reference \cite{2510}, the same energy-momentum tensor is recovered whether one assumes ${\cal L}_{matter}\to p$ and ${\cal L}_{matter}\to -\rho$.

The variation of the action with respct to the metric $g^{\mu\nu}$ leads to the following field equations
\begin{eqnarray}
&&A(\phi_c)\left[D_R R_{\mu \nu} +\left(P_T +\frac{1}{2}L_{{\cal
L}_{matter}}\right)\left(g_{\mu \nu} L_{matter}- T_{\mu
\nu}\right)-\frac{1}{2}g_{\mu \nu}F(R,T,{\cal L}_{matter}) \right.
\nonumber\\
&&\left. +\nabla_\mu \nabla_\nu D_R  - g_{\mu \nu}\Box D_R  \right] + g_{\mu \nu} \Lambda \nonumber\\
&& +\frac{1}{2}\left[g_{\mu \nu}
M^{ab}(\phi_c)G(\xi_{ab})+M^{ab}(\phi_c)\frac{\partial
G(\xi_{ab})}{\partial \xi_{ab}} \nabla_\mu \phi_a \nabla_\nu \phi_b
-g_{\mu \nu}C(\phi_c)\right]\nonumber\\
&&+D_R\nabla_\mu \nabla_\nu A(\phi_c) - g_{\mu \nu} D_R\,\Box
A(\phi) =-8\pi G_N T_{\mu \nu}, \label{einstein-frtl}
\end{eqnarray}
where  $D_R=\frac{\partial F(...)}{\partial R}$ and $P_T=\frac{\partial F(...)}{\partial T}$. These can be recast into an Einstein-like form as
\begin{eqnarray}
G_{\mu \nu}&=& \frac{1}{D_R}\left\{ g_{\mu \nu}\frac{\left[F(R,L) -
R\,D_R\right]}{2}- \nabla_\mu \nabla_\nu D_R  + g_{\mu \nu}\Box D_R
+\left(P_T +\frac{1}{2}L_{{\cal L}_{matter}}\right)\left(T_{\mu
\nu}-g_{\mu \nu}
L_{matter} \right)   \right\} \nonumber\\
&& - g_{\mu \nu} \frac{\Lambda}{A(\phi)D_R}
 -\frac{1}{2A(\phi_c)D_R}\left[g_{\mu \nu}
M^{ab}(\phi_c)G(\xi_{ab})+M^{ab}(\phi_c)\frac{\partial
G(\xi_{ab})}{\partial \xi_{ab}} \nabla_\mu \phi_a \nabla_\nu \phi_b
-g_{\mu \nu}C(\phi_c)\right]\nonumber\\
&&-\frac{1}{A(\phi_c)}\left[\nabla_\mu \nabla_\nu A(\phi_c) -g_{\mu
\nu} \Box A(\phi)\right] -\frac{8\pi G_N}{A(\phi_c)D_R} T_{\mu \nu}.
\label{einstein-like-frtl}
\end{eqnarray}
The variation of the action with respect to the scalar field $\phi_c$ yields the evolution equation
\begin{equation}
\frac{\partial A(\phi_j)}{\partial \phi_c} F(R,T,L)-\frac{\partial
M^{ab}(\phi_j)}{\partial
\phi_c}\,G(\xi_{ab})-M^{cb}(\phi_j)\frac{\partial
G(\xi_{ab})}{\partial \xi_{ab}}\,\Box \phi_b+\frac{\partial
C(\phi_j)}{\partial \phi_c}=0. \label{escalar-final}
\end{equation}
By excluding the matter Lagrangian from the functional F, the terms involving $L_{{\cal L}_{matter}}$ vanish. The field equations simplify accordingly to
\begin{eqnarray}
G_{\mu \nu}&=& \frac{1}{D_R}\left\{ g_{\mu \nu}\frac{\left[F(R,L) -
R\,D_R\right]}{2}- \nabla_\mu \nabla_\nu D_R  + g_{\mu \nu}\Box D_R
+P_T \left(T_{\mu \nu}-g_{\mu \nu}
L_{matter} \right)   \right\}  - g_{\mu \nu} \frac{\Lambda}{A(\phi)D_R} \nonumber\\
&& -\frac{1}{2A(\phi_c)D}\left[g_{\mu \nu}
M^{ab}(\phi_c)G(\xi_{ab})+M^{ab}(\phi_c)\frac{\partial
G(\xi_{ab})}{\partial \xi_{ab}} \nabla_\mu \phi_a \nabla_\nu \phi_b
-g_{\mu \nu}C(\phi_c)\right]\nonumber\\
&&-\frac{1}{A(\phi_c)}\left[\nabla_\mu \nabla_\nu A(\phi_c) -g_{\mu
\nu} \Box A(\phi)\right] -\frac{8\pi G_N}{A(\phi_c)D_R} T_{\mu \nu}.
\label{einstein-like-frt}
\end{eqnarray}
If the trace T is omitted from the functional, the auxiliary function $P_T$ is eliminated, resulting in
\begin{eqnarray}
G_{\mu \nu}&=& \frac{1}{D_R}\left\{ g_{\mu \nu}\frac{\left[F(R) -
R\,D_R\right]}{2}- \nabla_\mu \nabla_\nu D_R  + g_{\mu \nu}\Box D_R  \right\}  - g_{\mu \nu} \frac{\Lambda}{A(\phi)D_R} \nonumber\\
&& -\frac{1}{2A(\phi_c)D_R}\left[g_{\mu \nu}
M^{ab}(\phi_c)G(\xi_{ab})+M^{ab}(\phi_c)\frac{\partial
G(\xi_{ab})}{\partial \xi_{ab}} \nabla_\mu \phi_a \nabla_\nu \phi_b
-g_{\mu \nu}C(\phi_c)\right]\nonumber\\
&&-\frac{1}{A(\phi_c)}\left[\nabla_\mu \nabla_\nu A(\phi_c) -g_{\mu
\nu} \Box A(\phi)\right] -\frac{8\pi G_N}{A(\phi_c)D_R} T_{\mu \nu}.
\label{einstein-like-fr}
\end{eqnarray}
In all aforementioned instances, the scalar field equation remains invariant.
\subsubsection{Bergmann-Wagoner theory}
To recover the Bergmann-Wagoner and Brans-Dicke theories from this generalized framework, we apply the following identifications \cite{Clifford}
\begin{equation}
A(\phi)=\phi, \qquad M^{ab}(\phi_c)=-2\frac{\omega(\phi)}{\phi},
\qquad F(R)=R, \qquad C(\phi)=2\phi\, \lambda(\phi).
\end{equation}
Substituting these into the general equations, the scalar field equation becomes
\begin{equation}
\Box \phi + \frac{1}{2}\nabla_\mu \phi \nabla^\mu \phi
\frac{d}{d\phi}\left(Ln \left[\frac{\omega(\phi)}{\phi}\right]
\right)+\frac{1}{2}\frac{\phi}{\omega(\phi)}\left[R+2\,\frac{d}{d\phi}\left(\phi\,
\lambda(\phi) \right) \right]=0,
\end{equation}
which is consistent with Equation (5.32) in \cite{Clifford}. Similarly, the Einstein field equations reduce to
\begin{eqnarray}
R_{\mu \nu}-\frac{1}{2}R g_{\mu \nu}-\lambda(\phi) g_{\mu \nu}&=&
8\pi \phi^{-1} T_{\mu \nu}+ \phi^{-2}\omega(\phi)\left[\nabla_\mu
\phi \nabla_\nu \phi-\frac{1}{2}g_{\mu \nu} \nabla_\alpha \phi
\nabla^\alpha \phi
\right]\nonumber\\
&&+ \phi^{-1}\left[\nabla_\mu \nabla_\nu \phi- g_{\mu \nu} \Box \phi
\right]
\end{eqnarray}
reproducing Equation (5.31) in \cite{Clifford}.

\section{Exact Bianchi type I cosmological model with standard matter in the F(R) theory.}\label{section3}
In this section, we derive exact solutions for a Bianchi type I cosmological model within the framework of F(R) gravity. We consider the presence of standard matter modeled as a barotropic fluid with the equation of state $\rm P=\gamma \rho$. The barotropic parameter $\gamma$ assumes the values $\gamma <0,\frac{1}{3},1$ and $0$, corresponding to the inflation-like, radiation-dominated, stiff matter, and dust-dominated eras of cosmic evolution, respectively. The derivation is performed utilizing the Hamiltonian formalism.

\subsection{Barotropic fluid in standard matter}
We begin considering the gravitational action involving a modified Ricci scalar and a matter Lagrangian
\begin{equation}
\rm S=\int \sqrt{-g}\left[ F(R) + {\cal L}_{matter}\right]d^4x.\label{matter}
\end{equation}
To simplify the field equations, we introduce the auxiliary variable $\rm D=D_R=\frac{\partial F}{\partial R}$. In this representation, the energy-momentum tensor for standard matter is treated as a perfect fluid. Adjusting the general F(R) field equations for this specific model yields
\begin{eqnarray}
\rm D R_{\mu \nu} -\frac{1}{2}g_{\mu \nu}F(R) + \nabla_\mu \nabla_\nu D - g_{\mu \nu}\Box D  =-8\pi G T_{\mu \nu}. \label{einstein-i}
\end{eqnarray}
By taking the trace of the above expression, the structural constraint for F(R) theories is obtained
\begin{equation}
\rm F(R)=\frac{D R-3 \Box D+8\pi G T+}{2}. \label{fr}
\end{equation}
where T denotes the trace of the energy-momentum tensor $T_{\mu \nu}$.

The line element for the Bianchi type I model is expressed in terms of the lapse function $N$ and directional scale factors $A(t), B(t)$, and $C(t)$
\begin{equation}
ds^2=-N^2 dt^2 + A(t)^2dx^2+ B(t)^2dy^2 +C(t)^2dz^2, \label{line-normal}
\end{equation}
under the cosmic time transformation $Ndt=d\tau$, the metric simplifies to
\begin{equation}
ds^2=-d\tau^2 + A(\tau)^2dx^2+ B(\tau)^2dy^2 +C(\tau)^2dz^2. \label{line-conformal}
\end{equation}
The Ricci scalar R associated to the metrics (\ref{line-normal}) and (\ref{line-conformal}) are given by
 \begin{eqnarray}
R&=& \frac{\ddot A}{N^2A} +  \frac{\ddot B}{N^2B} + \frac{\ddot
C}{N^2C} + \frac{\dot A}{NA} \frac{\dot B}{NB} + \frac{\dot A}{NA}
\frac{\dot C}{NC} +\frac{\dot B}{NB}
\frac{\dot C}{NC} \nonumber\\
&&-\frac{\dot N}{N}\left[\frac{\dot A}{N^2 A}+ \frac{\dot B}{N^2
B}+\frac{\dot C}{N^2 C}\right], \label{ricci-normal}\\
R&=& \frac{A^{\prime \prime}}{A} +  \frac{B^{\prime \prime}}{B} +
\frac{C^{\prime \prime}}{C} + \frac{A^{\prime}}{A}
\frac{B^{\prime}}{B} + \frac{A^{\prime}}{A} \frac{C^{\prime}}{C}
+\frac{B^{\prime}}{B} \frac{C^{\prime}}{C}, \label{ricci}
\end{eqnarray}
where a dot denotes differentiation with respect to $t$, while a prime denotes differentiation with respect to $\tau$. It is relatively easy to see that in gauge N=1, the Ricci scalars, given by (\ref{ricci-normal}) and (\ref{ricci}), are equal.

The volume $\eta^3$ and generalized mean Hubble parameter H are defined as
\begin{equation}
\eta^3=V=ABC, \qquad H=\frac{1}{3}\left[H_1+H_2+H_3
\right]=\frac{1}{3}\frac{ V^{\prime}}{V},
\end{equation}
where $H_1=\frac{A^{\prime}}{A}, H_2=\frac{ B^{\prime}}{B}, H_3=\frac{
C^{\prime}}{C}$ are the directional Hubble parameters in the spatial directions (x, y, z), respectively. However, since $\frac{
V^{\prime}}{V}=3\frac{ \eta^{\prime}}{\eta}$, it follows that the Hubble parameter can be written as $H=\frac{ \eta^{\prime}}{\eta}$. In this framework, the Hubble parameter is expressed as a function of $\eta$ in the anisotropic cosmological model. In the standard Friedmann–Robertson–Walker metric (FRW) case, the deceleration parameter is defined in terms of the scale factor $a(t)$ as $q=-\frac{a\, \ddot a
}{{\dot a}^2}$; by analogy, we may introduce the deceleration parameter for the present model as a function of $\eta$, namely $q=-\frac{\eta \ddot \eta}{{\dot\eta}^2}$.

The energy-momentum conservation law $T^{\mu \nu};\nu=0$ leads to the following expression for energy density
\begin{equation}
 \rho^{\prime} + (1+\gamma)\rho \left[\frac{ A^{\prime}}{A} +\frac{
B^{\prime}}{B}+\frac{ C^{\prime}}{C}\right]=0, \qquad \rho =M_\gamma
\eta^{-3(1+\gamma)}, \label{conservation}
\end{equation}

From (\ref{einstein-i}) we can obtain the Einstien field equations, resulting in system of coupled differential equations for the scale factors and the field D, which read
\begin{eqnarray}
&&-12\pi G(1+\gamma) \frac{(ABC)^{-\gamma}}{D}-\frac{A^{\prime
\prime}}{ A}-\frac{ B^{\prime \prime}}{B}-\frac{C^{\prime
\prime}}{C}-\frac{3}{2}\frac{D^{\prime
\prime}}{D}+\frac{A^{\prime}}{A} \frac{B^{\prime}}{B} +
\frac{A^{\prime}}{A} \frac{C^{\prime}}{C} +\frac{B^{\prime}}{B}
\frac{C^{\prime}}{C}
 + \nonumber\\
 && \qquad + \frac{1}{2}\frac{D^{\prime}}{D}\left(\frac{A^{\prime}}{A} + \frac{B^{\prime}}{B} +\frac{C^{\prime}}{C}
\right) = 0, \label{00}\\
&&4\pi G(1+\gamma) \frac{(ABC)^{-\gamma}}{D}-\frac{ A^{\prime
\prime}}{ A}+\frac{ B^{\prime \prime}}{ B}+\frac{C^{\prime
\prime}}{C}+\frac{1}{2}\frac{ D^{\prime
\prime}}{D}-\frac{A^{\prime}}{A} \frac{B^{\prime}}{B} -
\frac{A^{\prime}}{A} \frac{C^{\prime}}{C} +\frac{B^{\prime}}{B}
\frac{C^{\prime}}{C}
 + \nonumber\\
 &&\qquad +\frac{1}{2}\frac{D^{\prime}}{D}\left(-3\frac{A^{\prime}}{A} + \frac{B^{\prime}}{B} +\frac{C^{\prime}}{C}
\right) = 0, \label{11}\\
&&4\pi G(1+\gamma) \frac{(ABC)^{-\gamma}}{D}+\frac{A^{\prime
\prime}}{A}-\frac{B^{\prime \prime}}{ B}+\frac{ C^{\prime
\prime}}{C}+\frac{1}{2}\frac{ D^{\prime \prime}}{
D}-\frac{A^{\prime}}{A} \frac{B^{\prime}}{B} + \frac{A^{\prime}}{A}
\frac{C^{\prime}}{C} -\frac{B^{\prime}}{B} \frac{C^{\prime}}{C}
 + \nonumber\\
 &&\qquad +\frac{1}{2}\frac{D^{\prime}}{D}\left(\frac{A^{\prime}}{A} -
3\frac{B^{\prime}}{B} +\frac{C^{\prime}}{C}
\right) = 0, \label{22}\\
&&4\pi G(1+\gamma) \frac{(ABC)^{-\gamma}}{D}+\frac{A^{\prime
\prime}}{ A}+\frac{ B^{\prime \prime}}{ B}-\frac{C^{\prime
\prime}}{C}+\frac{1}{2}\frac{D^{\prime
\prime}}{D}+\frac{A^{\prime}}{A} \frac{B^{\prime}}{B} -
\frac{A^{\prime}}{A} \frac{C^{\prime}}{C} -\frac{B^{\prime}}{B}
\frac{C^{\prime}}{C}
 + \nonumber\\
  && \qquad +\frac{1}{2}\frac{D^{\prime}}{D}\left(\frac{A^{\prime}}{A} + \frac{B^{\prime}}{B} -3\frac{C^{\prime}}{C}
\right) = 0, \label{33}
\end{eqnarray}
where (\ref{00}) corresponds to the temporal component, while (\ref{11})-(\ref{33}) correspond to the spatial components. By considering the differences between the spatial field equations, we isolate the evolution of the ratios between scale factors, in particular, subtracting equation (\ref{11}) from (\ref{22}), (\ref{33})
from (\ref{22}), and (\ref{11}) from (\ref{33}), we get respectively
\begin{eqnarray}
\frac{A^{\prime \prime}}{A}- \frac{B^{\prime \prime}}{B}+
\frac{C^{\prime}}{C}\left(\frac{A^{\prime}}{A}
-\frac{B^{\prime}}{B}\right)+\frac{D^{\prime}}{D}
\left(\frac{A^{\prime}}{A} - \frac{B^{\prime}}{B} \right) &=&0,
\label{11-0} \\
\frac{\ddot B^{\prime \prime}}{B}- \frac{C^{\prime \prime}}{C}+
\frac{A^{\prime}}{A}\left(\frac{B^{\prime}}{B}
-\frac{C^{\prime}}{C}\right)+\frac{D^{\prime}}{D}
\left(\frac{B^{\prime}}{B} - \frac{C^{\prime}}{C} \right) &=&0,
\label{22-0} \\
\frac{C^{\prime \prime}}{C}- \frac{A^{\prime \prime}}{A}+
\frac{B^{\prime}}{B}\left(\frac{C^{\prime}}{C}
-\frac{A^{\prime}}{A}\right)+\frac{D^{\prime}}{D}
\left(\frac{C^{\prime}}{C} - \frac{A^{\prime}}{A} \right) &=&0.
\label{33-0}
\end{eqnarray}
Equation (\ref{11-0}) can be rearrange as
\begin{equation}
BA^{\prime \prime} - A B^{\prime \prime} +
\left(\frac{C^\prime}{C}+\frac{D^\prime}{D} \right) \left(BA^\prime
-AB^\prime \right)=0, \label{modi1}
\end{equation}
however $\left(BA^\prime -AB^\prime \right)^\prime=BA^{\prime
\prime} - A B^{\prime \prime} $, then (\ref{modi1}) becomes
\begin{equation}
\frac{\left(BA^\prime -AB^\prime \right)^\prime}{\left(BA^\prime
-AB^\prime \right)} + \frac{d}{d\tau}Ln\left(CD \right) =0, \qquad
\to \qquad \frac{d}{d\tau}Ln\left[(BA^\prime-AB^\prime)CD \right]=0
\label{modi2}
\end{equation}
then
$(BA^\prime-AB^\prime)CD=\frac{d}{d\tau}Ln\left(\frac{A}{B}\right)ABCD=\frac{d}{d\tau}Ln\left(\frac{A}{B}\right)\eta^3
D=\alpha_1$, where $\alpha_1$ is a constant, and $\eta^3=ABC$. Integrating the last equation yields the ratio of the scle factors A and B
\begin{equation}
\frac{A}{B}=a_1 Exp\left[\alpha_1 \int \frac{d\tau}{\eta^3\,
D}\right], \label{ab}
\end{equation}
with $a_1$ an integration constant. Applying a similar procedure to the remaining directions take the form
\begin{eqnarray}
\frac{B}{C}&=& a_2\, Exp \left[\alpha_2 \int \frac{d\tau}{\eta^3\,
D}
 \right],  \label{bc}\\
 \frac{C}{A}&=& a_3\, Exp \left[\alpha_3 \int \frac{d\tau}{\eta^3\, D}
 \right],  \label{ac}
 \end{eqnarray}
where $a_i$ and $\alpha_i$ ($i=1,2,3$) are integration constants, which satisfy the relations 
 \begin{equation}
 \sum_{i=1}^3 \alpha_i=0, \qquad \Pi_{i=1}^3 a_i=1.
 \end{equation}
The general solutions for the scale factors expressed in quadrature form are
\begin{equation}
\rm A= v_1\, \eta Exp\left[u_1\int \frac{d\tau}{\eta^3\, D} \right],
\quad B=v_2 \,\eta Exp\left[u_2\int \frac{d\tau}{\eta^3\, D}
\right], \quad C=v_3\, \eta Exp\left[u_3\int \frac{d\tau}{\eta^3\,
D} \right], \label{factors}
\end{equation}
where the new constants are
\begin{equation}
\rm
v_1=\sqrt[3]{\frac{a_1}{a_3}},\quad v_2= \sqrt[3]{\frac{a_2}{a_1}},\quad v_3=\sqrt[3]{\frac{a_3}{a_2}} \quad u_1=\frac{\alpha_1-\alpha_3}{3}
\quad u_2= \frac{\alpha_2-\alpha_1}{3}, \quad
u_3=\frac{\alpha_3-\alpha_2}{3},
\end{equation}
we can note that these new constants satisfy the following relation
\begin{equation}
\rm \Pi_{i=1}^3 v_i=1, \qquad \sum_{i=1}^3 u_i=0.
\end{equation}
These solutions are closed once the functional forms of $\eta(\tau)$ and $D(\tau)$ are determined. In the subsequent section, we employ the Hamiltonian formalism to derive the specific conditions under which common power-law ansatzes, such as $D = \eta^m$ and $H \ell\eta^{-n}$  \cite{Uddin,Singh1}, emerge naturally.

\section{Lagrangian and Hamiltonian density }\label{section4}
The Lagrangian density ${\cal L}_I$, as defined in Eq. (\ref{matter}), is derived using the line element (\ref{line-normal}) and the Ricci scalar expression (\ref{ricci-normal}). The resulting relation is
\begin{eqnarray}
{\cal L}_I&=&-\frac{\ddot A}{N}BCD- \frac{\ddot B}{N}ACD
-\frac{\ddot C}{N}ABD -\frac{3}{2}\frac{\ddot D}{N}ABC -\frac{\dot A
\dot B}{N}
CD -\frac{\dot A \dot C}{N} BD - \frac{\dot B \dot C}{N} AD \nonumber \\
&&- \frac{3}{2}\frac{\dot D}{N}\left[\dot A BC +A \dot B C + AB \dot
C \right] + \frac{\dot N}{N^2}\left[\dot A BCD + A \dot B CD + AB
\dot C D +\frac{3}{2} ABC \dot D \right]+\nonumber\\
 && + 12\pi G N ABC (1+\gamma) \rho,\label{lagra-i}
\end{eqnarray}
by eliminating second-order time derivatives and incorporating the conservation law result (\ref{conservation}), the Lagrangian density simplifies to
\begin{eqnarray}
{\cal L}_I&=& \frac{\dot A \dot B}{N} CD +\frac{\dot A \dot C}{N} BD
+ \frac{\dot B \dot C}{N} AD + \frac{\dot D}{N}\left[\dot A BC +A
\dot B C + AB \dot C \right] + \nonumber\\
&& + 12\pi G N M_\gamma (1+\gamma) \eta^{-3\gamma}.
\label{lagra-end}
\end{eqnarray}
To derive the Hamiltonian density, we express the Lagrangian in canonical form, i.e ${\cal L}=\Pi_{q^\mu}\dot q^{\mu}-N {\cal H}$, where the lapse function N serves as a Lagrange multiplier. The variation with respect to N yields the primary constraint ${\cal H}=0$. For the coordinates $q^\mu=(A,B,C,D)$, the canonical momenta $\Pi_{q^{\mu}}=\frac{\partial {\cal L}}{\partial \dot q^{\mu}}$ and the corresponding velocities are determined as follows
\begin{eqnarray}
\Pi_A&=&\frac{\dot B}{N}CD+B\frac{\dot C}{N}D + BC\frac{\dot D}{N},
\qquad \dot A=\frac{N}{6\eta^3 D}\left(AB\Pi_B +
AC\Pi_C+AD\Pi_D-2A^2 \Pi_A \right), \label{pa}\\
\Pi_B&=&\frac{\dot A}{N}CD+A\frac{\dot C}{N}D + AC\frac{\dot D}{N},
\qquad \dot B=\frac{N}{6\eta^3 D}\left(AB\Pi_A +
BC\Pi_C+BD\Pi_D-2B^2 \Pi_B \right), \label{pb}\\
\Pi_C&=&\frac{\dot A}{N}BD+A\frac{\dot B}{N}D + AB\frac{\dot D}{N},
\qquad \dot C=\frac{N}{6\eta^3 D}\left(AC\Pi_A +
BC\Pi_B+CD\Pi_D-2C^2 \Pi_C \right), \label{pc}\\
\Pi_D&=&\frac{\dot A}{N}BC+A\frac{\dot B}{N}C + AB\frac{\dot C}{N},
\qquad \dot D=\frac{N}{6\eta^3 D}\left(AD\Pi_A +
BD\Pi_B+CD\Pi_C-2D^2 \Pi_D \right). \label{pd}
\end{eqnarray}
Substituting these into the Lagrangian yields the general Hamiltonian density (in any gauge N)
\begin{eqnarray}
{\cal H}&=& \frac{1}{6\eta^3 D}\left[ -A^2 \Pi_A^2-B^2 \Pi_B^2-C^2
\Pi_C^2-D^2 \Pi_D^2 +A\Pi_A\, B\Pi_B + A\Pi_A\, C\Pi_C + B\Pi_B
C\Pi_C \right. \nonumber\\
&& \left. +D\Pi_D\left( A\Pi_A + B\Pi_B+C\Pi_C\right) -72\pi G
M_\gamma (1+\gamma) \eta^{3(1-\gamma)} D\right]. \label{hami}
\end{eqnarray}
To simplify the dynamics, we employ the transformation $Q=e^q$. Utilizing the Hamilton-Jacobi relation $\Pi_Q=\frac{\partial S}{\partial Q}=\frac{\partial S}{\partial q}
\frac{\partial q}{\partial Q}$, we define the transformed momenta $P_q=\frac{\partial S}{\partial q}=Q\Pi_Q$. The Hamiltonian density in terms of logarithmic variables $(a,b,c,d)$ becomes
\begin{eqnarray}
{\cal H}&=& \frac{e^{-a-b-c-d}}{6}\left[ - P_a^2- P_b^2- P_c^2-
P_d^2 +P_a P_b + P_a P_c + P_b P_c  +P_d\left( P_a + P_b+P_c\right)
\right.-\nonumber\\
&&\left. -72\pi G M_\gamma(1+\gamma) e^{(1-\gamma)(a+b+c)+d}
\right]. \label{hami-n}
\end{eqnarray}
The resulting Hamilton equations $ q^{\prime}=\frac{\partial {\cal H}}{\partial P_q}$ and $P^{\prime}_q=-\frac{\partial {\cal H}}{\partial q}$ are
\begin{eqnarray}
 a^{\prime}&=& \frac{1}{6\eta^3 D}\left[-2P_a+P_b+P_c+P_d
\right],\label{dai}\\
 b^{\prime}&=& \frac{1}{6\eta^3 D}\left[-2P_b+P_a+P_c+P_d
\right],\label{dbi}\\
 c^{\prime}&=& \frac{1}{6\eta^3 D}\left[-2P_c+P_a+P_b+P_d
\right],\label{dci}\\
 d^{\prime}&=&\frac{1}{6\eta^3 D}\left[-2P_d+P_a+P_b+P_c
\right],\label{ddi}\\
 P^{\prime}_i &=& \frac{1}{6\eta^3 D} \left[ 72\pi G M_\gamma (1-\gamma^2)
e^{(1-\gamma)(a+b+c)+d}\right], \label{pabci}\       i=\{a,b,c\}\\
 P^{\prime}_d &=& \frac{1}{6\eta^3 D} \left[ 72\pi G M_\gamma (1+\gamma) e^{(1-\gamma)(a+b+c)+d}\right].
\label{pdi}
\end{eqnarray}
By summing the equations for a $a^{\prime}$, $b^{\prime}$ and $c^{\prime}$  we identify the Hubble parameter relation
\begin{eqnarray}
H=\frac{\eta^{\prime}}{\eta}=\frac{P_d}{6\eta^3 D} \label{hub-pd},
\end{eqnarray}
from which we obtain
\begin{eqnarray}
    \eta&=&Exp\left(\int \frac{P_d}{6\eta^3 D} d\tau\right) \label{eta-pd}.
\end{eqnarray}
From the momentum evolution equations (\ref{pabci}) and (\ref{pdi}), we deduce that
\begin{eqnarray}
P_i&=&(1-\gamma)P_d+\alpha_i, \ \ \ i={a,b,c}
\end{eqnarray}
and integrating the coordinate equations yields the general solutions for the scale factors
\begin{eqnarray}
    A &=& \beta_a \eta Exp \left[ (-2\alpha_a+\alpha_b+\alpha_c) \int \frac{d\tau}{6\eta^3 D} \right] \label{a-int}, \\
    B &=& \beta_b \eta Exp \left[ (-2\alpha_b+\alpha_a+\alpha_c) \int \frac{d\tau}{6\eta^3 D} \right] \label{b-int},\\
    C &=& \beta_c \eta Exp \left[ (-2\alpha_c+\alpha_a+\alpha_b) \int \frac{d\tau}{6\eta^3 D} \right] \label{c-int},\\
    D &=& \beta_d \eta^{1-3\gamma} Exp \left[ (\alpha_a+\alpha_b+\alpha_c) \int \frac{d\tau}{6\eta^3 D} \right]. \label{d-int}
\end{eqnarray}
The final set of equations is consistent with (\ref{factors}); however, an additional fourth equation is derived. From this, we identify the standard potential form $D\propto \eta^{m}$, which arises under the condition $\alpha_a+\alpha_b+\alpha_c=0$. Notably, the expressions for the scale factors A, B, and C remain in accordance with the Einstein field equations.
\subsection{Solutions in $D \propto \eta^{m}$ and $H \propto \eta^{-n}$ ansatz}
In many contexts, $D$ and $H$ are assume to follow power-law behaviors, namely $D \propto \eta^{m}$ and $H \propto \eta^{-n}$; however, the values of the constants $m,n$ are not properly justified. In this section, a physical meaning will be given to them, and the corresponding dynamical equations will be written. As discussed above, the choice $\alpha_a+\alpha_b+\alpha_c=0$ leaves the functions $A, B, C$ unchanged; that is, they continue to satisfy Einstein field equations, namely
\begin{eqnarray}
    A &=& \beta_a \eta Exp \left[ -3\alpha_a \int \frac{d\tau}{6\eta^3 D} \right] \label{daf},\\
    B &=& \beta_b \eta Exp \left[ -3\alpha_b \int \frac{d\tau}{6\eta^3 D} \right] \label{dbf},\\
    C &=& \beta_c \eta Exp \left[ -3\alpha_c \int \frac{d\tau}{6\eta^3 D} \right] \label{dcf},\\
    D &=& \beta_d \eta^{1-3\gamma},
\end{eqnarray}
we can notice that $D = \beta_d \eta^{1-3\gamma}$ (identifying $m = 1-3\gamma$). For an ansatz $H = \ell \eta^{-n}$, together with Eq. (\ref{hub-pd}), the variable $P_d$ is written as
\begin{eqnarray}
P_d&=&6\beta_d\ell \eta^{4-n-3\gamma},
\end{eqnarray}
where for consistency, according to field equation (\ref{pdi}), it is required that 
\begin{eqnarray}
n&=&2 ,\\
\ell&=&\sqrt{\frac{2 \pi G M_\gamma (1+\gamma)}{\beta_d
(2-3\gamma)}}. \label{ll}
\end{eqnarray}
Its important to emphasize that the above result is valid for values of $\gamma$ within the open interval $(-1,+\frac{2}{3})$. Integrating equation (\ref{hub-pd}), rewritten as $\frac{\eta^{\prime}}{\eta}=\ell \eta^{-2}$, gives us the temporal evolution
\begin{eqnarray}
\eta&=&\left( 2\ell \tau +\kappa_0 \right)^{1/2}, \label{volumen-d} \\
D&=&\beta_d \left( 2\ell \tau +\kappa_0
\right)^{(1-3\gamma)/2},\label{auxiliary}
 \end{eqnarray}
 where $\kappa_0=-2\ell \tau_0+\eta^{2}_0$. The specific scale factors $A, B, C$ are then expressed as functions of proper time $\tau$, illustrating how matter-energy content governs the universe's expansion
\begin{eqnarray}
    A &=& \beta_a \left( 2\ell \tau +\kappa_0 \right)^{1/2} Exp \left[ \frac{3 \alpha_a}{6 \beta_d \ell (2-3\gamma)} \left[  \left( 2\ell \tau +\kappa_0 \right)^{(3\gamma-2)/2}-\eta^{3\gamma-2}_0 \right]  \right], \\
    B &=& \beta_b \left( 2\ell \tau +\kappa_0 \right)^{1/2} Exp \left[ \frac{3 \alpha_b}{6 \beta_d \ell (2-3\gamma)} \left[  \left( 2\ell \tau +\kappa_0 \right)^{(3\gamma-2)/2}-\eta^{3\gamma-2}_0 \right]  \right], \\
    C &=& \beta_c \left( 2\ell \tau +\kappa_0 \right)^{1/2} Exp \left[ \frac{3 \alpha_c}{6 \beta_d \ell (2-3\gamma)} \left[  \left( 2\ell \tau +\kappa_0 \right)^{(3\gamma-2)/2}-\eta^{3\gamma-2}_0 \right]  \right], \\
    D&=&\beta_d \left( 2\ell \tau +\kappa_0 \right)^{(1-3\gamma)/2}.
\end{eqnarray}
Evidently, the values $\gamma=\frac{2}{3}$ and $\gamma=-1$ fall outside the scope of this approximation. 
While the subsequent section investigates the $\gamma=-1$ case—specifically regarding inflationary and empty universes—we first illustrate the graphical evolution of the volume function and the auxiliary variable $D=\frac{dF(R)}{dR}$, conceived as a type of "scalar field" with a geometric structure.

\begin{figure}[ht!]
\begin{center}
\captionsetup{width=0.9\textwidth}
\includegraphics[scale=0.43]{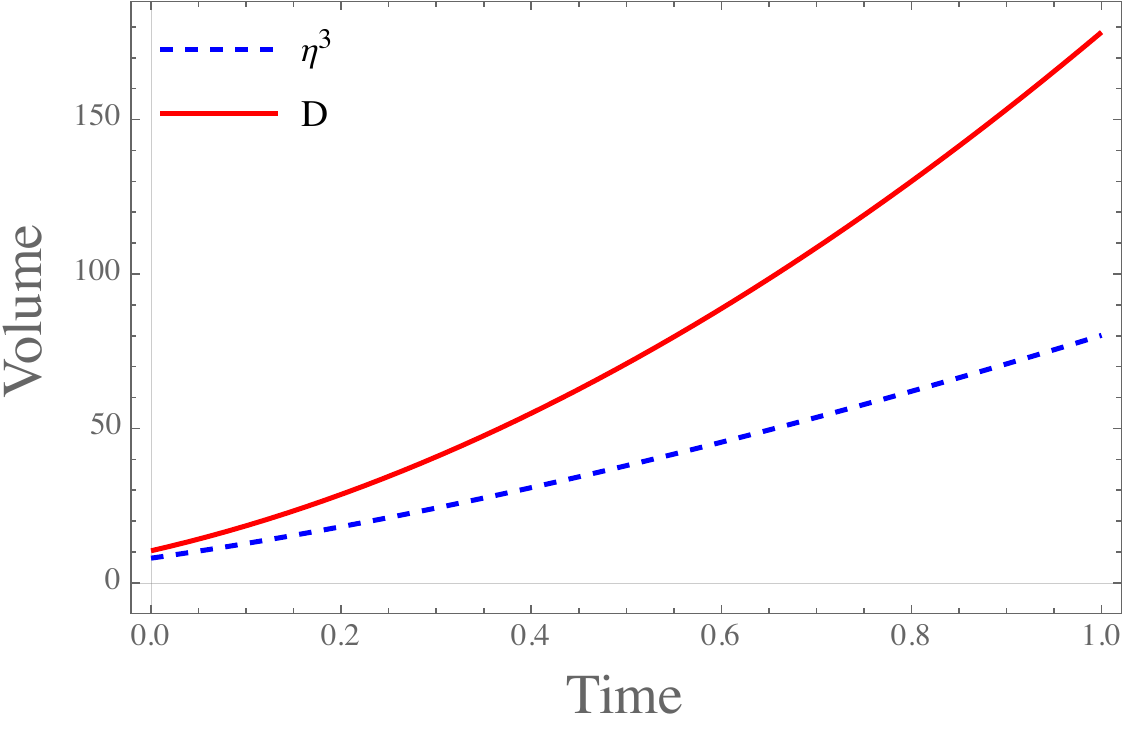}~~
\includegraphics[scale=0.43]{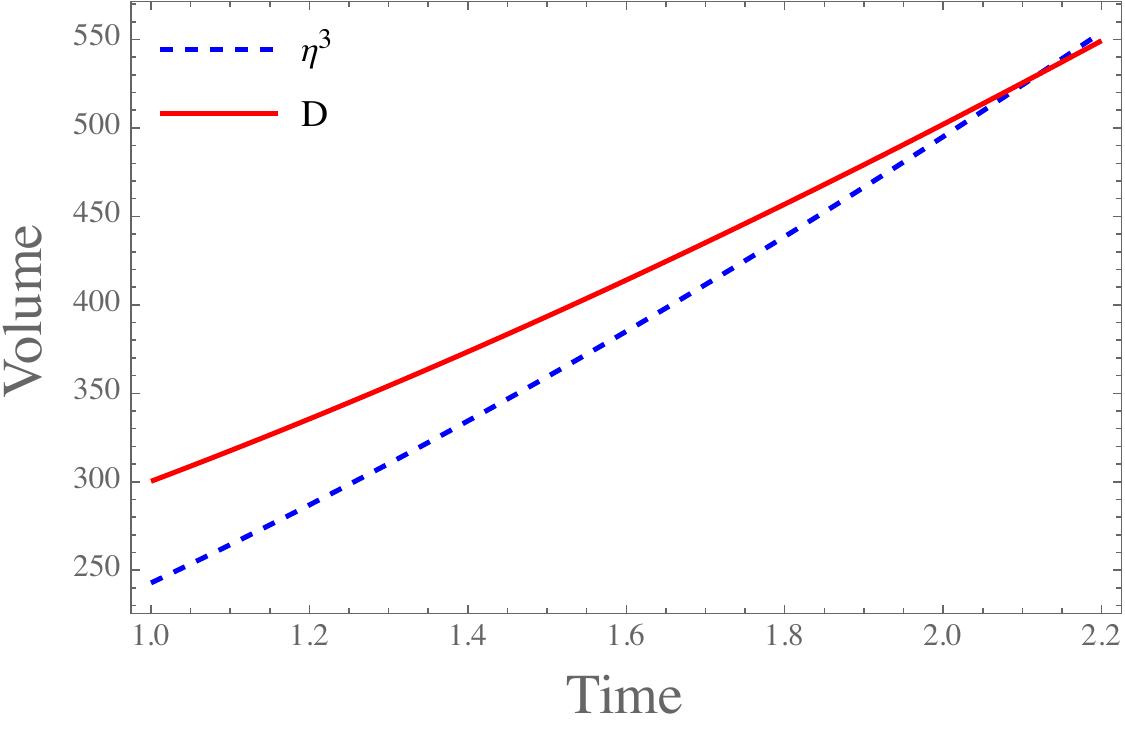}
\caption{In the left panel we shown the behavior of the volume
function and the auxiliary function D in the inflation like epoch
with $\gamma=-0.9$, $\kappa=4$ and $\beta_d=0.8$ and $8\pi G_N
M_{-1}=1000$, constant that appear in the definition of the $\ell$
parameter, equation (\ref{ll}). In the right panel, we consider the
post inflationary epoch with $\gamma=-\frac{2}{3}$ and $8\pi G_N
M_{-1}=800$, taking $\eta_{-0.9}=170$ and $\beta_{d_{-.09}}=80$.
\label{graf1}}
\end{center}
\end{figure}

Each plot simultaneously depicts the temporal evolution of the model’s volume alongside the behavior of the auxiliary function D. In figure \ref{graf1}, the parameter $\gamma$ is chosen to be very close to $-1$ to observe the evolution during the primordial inflationary epoch, assigning specific values to the parameters appearing in equations (\ref{volumen-d}) and (\ref{auxiliary}). Based on these results, we adopt the final value of each function (the volume and the auxiliary function D) as the initial condition for the subsequent epoch. The evolution intervals are selected arbitrarily; specifically, the second and following plots are defined by $\eta^3_\gamma=\eta_i +\,\eta^3$, where $\eta_i$ denotes the final value of each epoch $\gamma$, with a similar procedure applied to the auxiliary function D, such that $D_\gamma=\beta_{d_{i}}+ D(\tau)$.

Fig.~\ref{graf1} displays the first two scenarios: inflation where $\gamma=-0.9$ and post-inflation where $\gamma=-2/3$. Within the considered time interval, the evolution of the auxiliary function D remains above that of the volume function. This contrast is most evident in the first scenario, whereas in the second, the two curves intersect. Following this intersection, the volume function continues to increase while the auxiliary function remains constant during the dust stage. Notably, during the $\gamma=-2/3$ stage, the auxiliary function exhibits a linear behavior over time, remaining below the volume, as the latter consistently follows a $3/2$ power-law evolution. The process in which the auxiliary function D exceeds the volume function may represent the mechanism by which cosmic inflation ceases, allowing matter to reorganize before entering subsequent stages. Until this crossing occurs, evolution proceeds at a slower rate but continues indefinitely; the presence of this auxiliary function persists through the following cosmic epochs as a background.

\begin{figure}[ht!]
\begin{center}
\captionsetup{width=0.9\textwidth}
\includegraphics[scale=0.425]{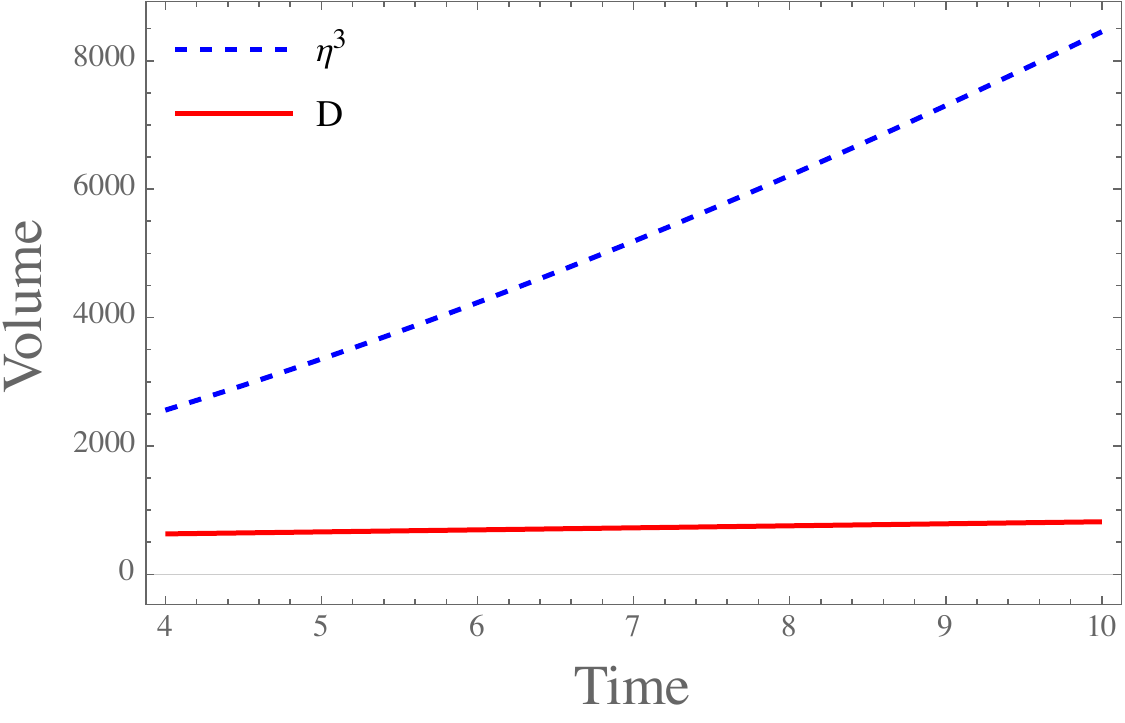}~~
\includegraphics[scale=0.425]{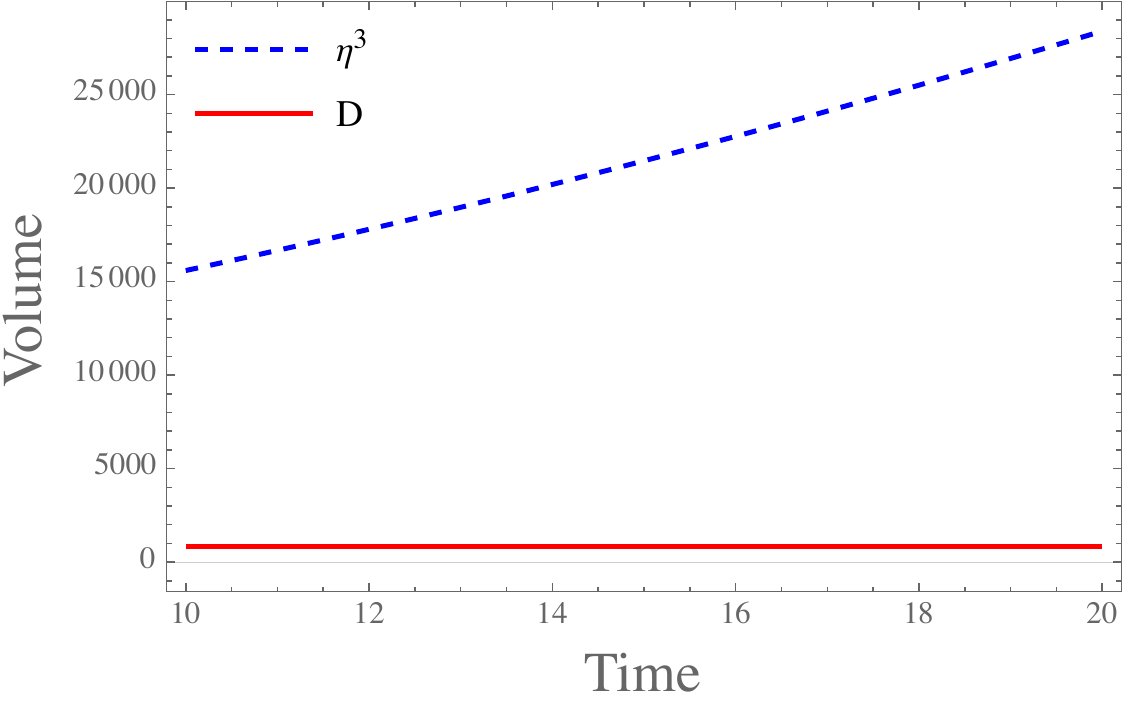}
\caption{The left panel shows the behavior of the volume
function and the auxiliary function D in the post-inflation like
epoch with $\gamma=-\frac{1}{3}$,  and $8\pi G_N M_{-1}=700$, taking
$\eta_{-\frac{2}{3}}=500$ and $\beta_{d_{\frac{2}{3}}}=500$. In the right panel, the radiation epoch is portrayed, where $\gamma=\frac{1}{3}$, $8\pi
G_N M_{-1}=100$, taking $\eta_{-0.9}=8500$ and
$\beta_{d_{-.09}}=850$. \label{graf2}}
\end{center}
\end{figure}

In figure \ref{graf2}, is depiected other post-inflationary cases such as $\gamma=-1/3$ and $\gamma=1/3$ (radiation), the volume function dominates the evolution of the auxiliary function D throughout the interval. In these instances, the volume function continues its growth while the auxiliary function remains constant during the matter stage, persisting as a cosmic background.

\begin{figure}[ht!]
\begin{center}
\captionsetup{width=0.9\textwidth}
\includegraphics[scale=0.5]{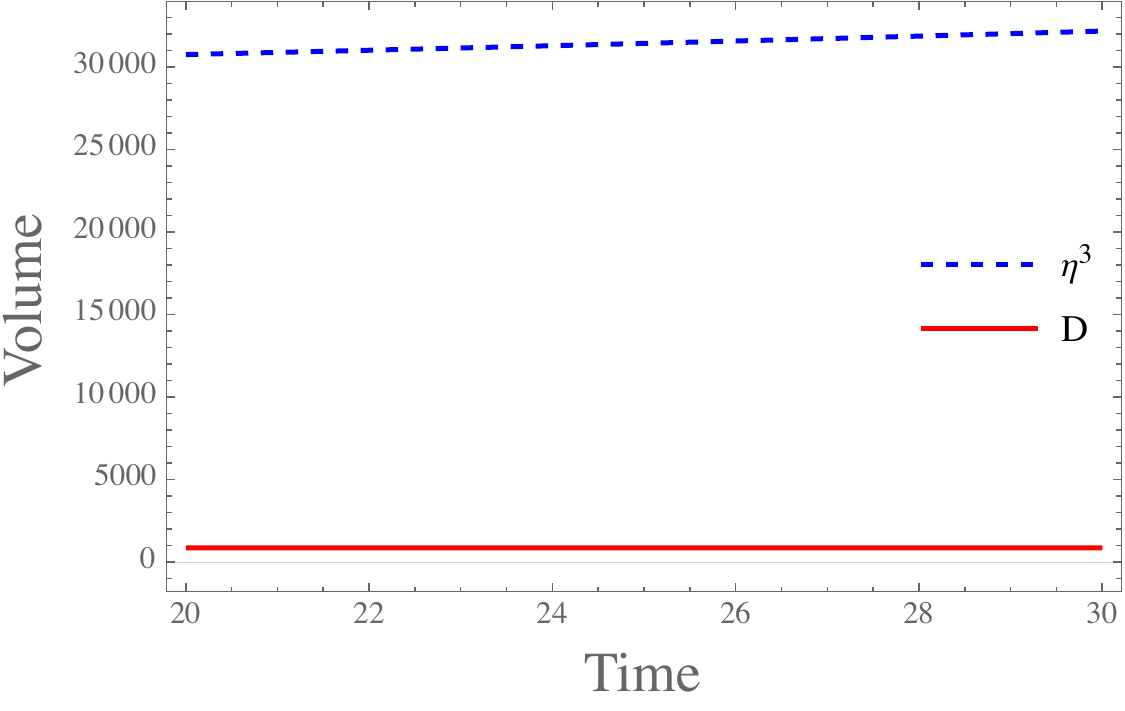}
\caption{The figure shows the bahavior of the volume and auxiliary D functions in the dust stage, where $\gamma=0$, $8\pi G_N M_{-1}=10$, $\eta_{\frac{1}{3}}=29000$ and $\beta_{d_{\frac{1}{3}}}=850$.
\label{graf3}}
\end{center}
\end{figure}

Figure \ref{graf3} displays the dust scenario, wherein the volume exhibits an accelerated behavior; nevertheless, the auxiliary function remains constant throughout this dust-dominated stage.
\subsection{Inflationary case $\gamma=-1$}
In the inflationary scenario, characterized by $\gamma=-1$ (equivalent to the vanishing matter density condition $M_\gamma=0$), the momenta $p_a, p_b, p_c, p_d$ are constant in time (due to the fact that $
P^{\prime}_a= P^{\prime}_b= P^{\prime}_c= P^{\prime}_d=0$). By summing the evolutionary equations for these variables, we obtain the relation for the product of the volume scale factor and the auxiliary field D:
\begin{eqnarray}
    6 \eta^3 D = \int \left[ P_a+P_b+P_c+P_d \right] d\tau=\kappa +p_0(\tau-\tau_0), \label{6vol}
\end{eqnarray}
where $p_0=\sum {p_i}$ with $i=a,b,c,d$. Integrating the coordinate equations (\ref{dai})-(\ref{ddi}), we obtain
\begin{eqnarray}
a&=&a_0 +Ln \left[\kappa + p_0(\tau-\tau_0)
\right]^{\frac{p_1}{p_0}}, \qquad p_1=-2p_a+p_b+p_c+p_d,
\label{aa-1}\\
b&=&b_0 + Ln \left[\kappa + p_0(\tau-\tau_0)
\right]^{\frac{p_2}{p_0}}, \qquad p_2=-2p_b+p_a+p_c+p_d,
\label{bb-1}\\
c&=&c_0 + Ln \left[\kappa + p_0(\tau-\tau_0)
\right]^{\frac{p_3}{p_0}}, \qquad p_3=-2p_c+p_a+p_b+p_d,
\label{cc-1}\\
d&=&d_0 + Ln \left[\kappa + p_0(\tau-\tau_0)
\right]^{\frac{p_4}{p_0}}, \qquad p_4=-2p_d+p_a+p_b+p_c,
\label{dd-1}
\end{eqnarray}
thus, the scale factors and the auxiliary function become
\begin{eqnarray}
A&=&A_0  \left[\kappa + p_0(\tau-\tau_0)
\right]^{\frac{p_1}{p_0}},
\label{aa-1}\\
B&=&B_0  \left[\kappa + p_0(\tau-\tau_0)
\right]^{\frac{p_2}{p_0}},
\label{bb-1}\\
C&=&C_0  \left[\kappa + p_0(\tau-\tau_0)
\right]^{\frac{p_3}{p_0}},
\label{cc-1}\\
D&=&D_0 \left[\kappa + p_0(\tau-\tau_0)
\right]^{\frac{p_4}{p_0}}.
\label{dd-1}
\end{eqnarray}
The generalized scale factor $\eta$ follows a power-law behavior of the form
\begin{equation}
\eta=\eta_0\left[\kappa + p_0(\tau-\tau_0)\right]^{\frac{p_d}{p_0}},
\end{equation}
from which, the volume function reads
\begin{equation}
\eta^3=\eta_0^3\left[\kappa + p_0(\tau-\tau_0)\right]^{\frac{p_1+p_2+p_3}{p_0}}=\eta_0^3\left[\kappa +p_0(\tau-\tau_0) \right]^{\frac{3p_d}{p_0}}. \label{volum}
\end{equation}
where, first we identified that $\eta^3D=\eta_0^3D_0 \left[\kappa +p_0(\tau-\tau_0) \right]$. To ensure an inflationary expansion we require that $\frac{3p_d}{p_0}>1$, and that the following constraint equation between the constants is satisfied
 \begin{equation}
 - p_a^2- p_b^2- p_c^2-p_d^2 +p_a p_b + p_a p_c + p_b p_c  +p_d\left( p_a +p_b+p_c\right)=0.
\end{equation}
Under these conditions, the auxiliary function D can be expressed as a power of the scale factor
\begin{equation}
D=d_0 \eta^m, \qquad{\rm with}~~m=\frac{p_4}{p_d}.
\end{equation}
This last result is consistent with the solutions derived in the previous section (equations (\ref{factors})). 

We can calculate the Hubble parameter
\begin{equation}
H=\frac{ \eta^{\prime}}{\eta}=\frac{p_d}{[\kappa +p_0(\tau-\tau_0)]},
\end{equation}
from which, the Ricci scalar (\ref{ricci}) takes the form 
\begin{equation}
R=\frac{\ell}{18\left(\kappa +p_0(\tau-\tau_0) \right)^2},.
\end{equation}
where the constant $\ell=p_0(p_1+p_2+p_3)-p_1(p_1+p_2+p_3)-p_2(p_2+p_3)-p_3^2$. From (\ref{fr}), the corresponding F(R) function is
\begin{equation}
\rm F(R(t))=\frac{d_0}{36}\left(\kappa+p_0(\tau-\tau_0)\right)^{\frac{p_4}{p_0}}\left[\frac{\ell}{\left(\kappa+p_0(\tau-\tau_0) \right)^2} \right], \label{frt}
\end{equation}
and upon further manipulation the functional form of F(R) can be expressed explicitly in terms of the Ricci scalar as
\begin{equation}
F(R)= \frac{d_0}{36}\left(\ell\right)^{\frac{p_4}{2p_0}} R^{1-\frac{p_4}{2p_0}}. \label{fr-s}
\end{equation}
However, for these solutions to satisfy the Einstein field equations (\ref{00})–(\ref{33}), the constraint $\ell=0$ must hold. This implies that both the Ricci scalar and the F(R) function vanish, effectively recovering the vacuum results of standard General Relativity.

\subsection{Hamiltonian in the gauge $N=6\eta^3 D$}

By adopting the specific gauge $N=6\eta^3 D$, where $d\tau=Ndt$, the Hamilton equations, (\ref{dai})-(\ref{pdi}), simplify significantly, giving
\begin{eqnarray}
\dot a&=& -2P_a+P_b+P_c+P_d,\label{da-n}\\
\dot b&=& -2P_b + P_a + P_c +P_d,\label{db-n}\\
\dot c&=& -2P_c + P_a +P_b +P_d,\label{dc-n}\\
\dot d&=& -2P_d + P_a +P_b +P_c,\label{dd-n}\\
\dot P_i &=&=72\pi G M_\gamma (1-\gamma^2)
e^{(1-\gamma)(a+b+c)+d}, \ \ i={a,b,c}
\label{pabc-n}\\
\dot P_d &=& 72\pi G M_\gamma (1+\gamma) e^{(1-\gamma)(a+b+c)+d}.
\label{pd-n}
\end{eqnarray}
We proceed to solve the Hamilton equations sectorally by employing the Hamiltonian constraint to obtain an ordinary differential equation for particular momenta, which are then mapped back to the temporal evolution of the coordinates $(a,b,c,d)$. Analysis of equations (\ref{pabc-n}) and (\ref{pd-n}) yields the relation $P_i=(1-\gamma)P_d+\alpha_i$, for $i={a,b,c}$. Substituting this expression along with Eq. (\ref{pd-n}) into Eq. (\ref{hami-n}) under the constraint ${\cal H}=0 $ results in the governing differential equation for $P_d$
\begin{eqnarray}
-P^{2}_d+b_0 P_d -c_0=\frac{\dot P_d}{3\gamma-2},\label{98}
\end{eqnarray}
where
\begin{eqnarray}
b_0&=&\frac{\alpha_a+\alpha_b+\alpha_c}{3\gamma-2},\\
c_0&=&\frac{\alpha_a^{2}+\alpha_b^{2}+\alpha_c^{2}-\alpha_a\alpha_b-\alpha_a\alpha_c-\alpha_b\alpha_c}{3\gamma-2}.
\end{eqnarray}
Equation (\ref{98}) can be recasted as
\begin{equation}
\frac{4dP_d}{\omega^2 -(2P_d -b_0)^2}=(3\gamma-2)dt, \qquad \omega=\sqrt{b_0^2-4c_0},
\end{equation}
whose solution is giving by
\begin{equation}
    P_d=\frac{b_0}{2}+\frac{\omega}{2}\, Tanh\left[\frac{\omega}{2}(3\gamma-2) (t-t_0)  \right].
\end{equation}
Now, we are in a position to calculate the generalized scale factor $\eta$ (given by equation (\ref{eta-pd}))
\begin{eqnarray}
    \eta&=&Exp\left(\frac{b_0}{2}(t-t_0)\right)\left(Cosh\left[\frac{\omega}{2}(3\gamma-2) (t-t_0)  \right] \right)^{1/(3\gamma-2)}.
\end{eqnarray}
The directional scale factors A,B,C and the function D are then obtained by integration
\begin{eqnarray}
A&=&\beta_a\eta_0 Exp\left[\left(\frac{b_0}{2}-2\alpha_a+\alpha_b+\alpha_c\right)(t-t_0)\right]\left(Cosh\left[\frac{\omega}{2}(3\gamma-2) (t-t_0)  \right] \right)^{1/(3\gamma-2)} \\
B&=&\beta_b\eta_0 Exp\left[\left(\frac{b_0}{2}-2\alpha_b+\alpha_a+\alpha_c\right)(t-t_0)\right]\left(Cosh\left[\frac{\omega}{2}(3\gamma-2) (t-t_0)  \right] \right)^{1/(3\gamma-2)} \\
C&=&\beta_c\eta_0 Exp\left[\left(\frac{b_0}{2}-2\alpha_c+\alpha_a+\alpha_b\right)(t-t_0)\right]\left(Cosh\left[\frac{\omega}{2}(3\gamma-2) (t-t_0)  \right] \right)^{1/(3\gamma-2)} \\
D&=&\beta_d\eta_0^{1-3\gamma}
Exp\left[\left((1-3\gamma)\frac{b_0}{2}+\alpha_a+\alpha_b+\alpha_c\right)(t-t_0)\right]\left(Cosh\left[\frac{\omega}{2}(3\gamma-2)
(t-t_0)  \right] \right)^{(1-3\gamma)/(3\gamma-2)}. \label{fun-d}
\end{eqnarray}
In the specific case of stiff matter epoch, $\gamma=1$, the solution for the auxiliary function D is derived directly from Eq. (\ref{pd-n}). By considering the time derivative of the momentum $P_d$, we obtain 
\begin{equation}
D=e^d=\frac{\dot P_d}{144\pi G M_1}=\frac{\omega^2}{576\pi G M_1}Sech^2\left[\frac{\omega (t -t_1)}{2} \right], \label{auxiliary-D}
\end{equation}
where the consistency of the constants requires the relation $\beta_d=\frac{\omega^2 \eta_0^2}{576 \pi G_N M_1}$. It should be noted that the preceding equations are not applicable to the cases where $\gamma=-1$ or $\gamma=2/3$.

\subsection{Inflation case, $\gamma=-1$ in $N=6\eta^3D$ gauge}
This scenario is equivalent to the case where $M_\gamma=0$. In this regime, all canonical momenta—denoted by $p_a, p_b, p_c$ and $p_d$—remain constant over time. Consequently, the coordinate functions are found to be
\begin{eqnarray}
a(t)&= a_1 + p_1 (t-t_{-1}), \qquad p_1=-2p_a+p_b+p_c+p_d, \label{an-n} \\
b(t)&= b_1 + p_2 (t-t_{-1}), \qquad p_2=-2p_b+p_a+p_c+p_d,\label{bn-n}\\
c(t)&= c_1 + p_3 (t-t_{-1}), \qquad p_3=-2p_c+p_a+p_b+p_d,\label{cn-n} \\
d(t)&= d_1 + p_4(t-t_{-1}),\qquad p_4=-2p_d+p_a+p_b+p_c,\label{dn-n}
\end{eqnarray}
where $p_0=p_a+p_b+p_c+p_d=p_1+p_2+p_3+p_4$. The corresponding directional scale factors and the auxiliary function D are expressed as
\begin{eqnarray}
A(t)&=& A_1 Exp\left[ p_1 (t-t_{-1})\right], \label{scale-a}\\
B(t)&=& B_1 Exp\left[ p_2 (t-t_{-1})\right], \label{scale-b}\\
C(t)&=& C_1 Exp\left[ p_3 (t-t_{-1})\right], \label{scale-c}\\
D(t)&=& D_1 Exp\left[ p_4 (t-t_{-1})\right], \label{scale-d}
\end{eqnarray}
with $Q_i=e^{q_i}=(A_1,B_1,C_1,D_1)$ serving as integration constants. Furthermore, the volume evolution is given by
\begin{eqnarray}
\eta^3\,D&=&\eta_0^2D_0 Exp\left[p_0 (t-t_{-1})\right]\quad \Rightarrow \quad
\eta^3 = \eta_0^3 e^{3p_d(t-t_{-1})}, \label{volume}
\end{eqnarray}
while the Ricci scalar takes the form
\begin{equation}
R=\frac{\ell}{18R_0^2}e^{2p_0(t-t_0)}, \label{ricci-n}
\end{equation}
where $\ell$ is a constraint between the constants $(p_a, p_b, p_c, p_d)$, and is given by $\ell=-p_a^2-p_b^2-p_c^2-p_d^2+p_a\,p_b+p_a\,p_c+p_a\,p_d+p_b\,p_c+p_b\,p_d+p_c\,p_d$. For these solutions to satisfy the Einstein field equations, the constants must obey the constraint $\ell=0$. Under this condition, the model yields a constant Hubble parameter $H=\frac{\dot \eta}{\eta}=p_d$ and a deceleration parameter $\rm q=-1$, characterizing an inflationary epoch. Notably, the Ricci scalar vanishes, consistent with standard General Relativity in a vacuum, which contradicts the temporal structures proposed in previous literature \cite{Sharif2009, Sharif2010}. Our analysis suggests that the solutions in those works do not satisfy the field equations under the stated premises.

\section{Conclusions}\label{section5}
In this work we employed a Hamiltonian framework to obtain exact solutions for the cosmic scale factors (A, B, C) and the auxiliary field $\rm D=\partial F/\partial R$. Through the application of specific consistency constraints, we recovered the conventional ansätze used in algebraic treatments of modified field equations. We demonstrate that the parameters of these ansätze are fundamentally coupled to the barotropic index $\gamma$, thereby linking the geometric evolution of the theory to the fluid properties of the corresponding cosmological era.

Although these results constitute a distinct class of classical solutions, the formal structure of the Hamiltonian density suggests the existence of broader solution sets under different gauge choices. We conclude that several prior solutions in the field remain incomplete, as they often overlook the rigorous constraints imposed by the full set of field equations. Extensions of this methodology to the quantum regime via the Wheeler-DeWitt equation and the exploration of more complex F(…) functionals are currently in progress.

\acknowledgments{ \noindent This work was partially supported by
PROMEP grants UGTO-CA-3. A.E.G, S.P.P, L.R.D.B and J.S. was
partially supported SNI-CONACYT. Many calculations where done by
Symbolic Program REDUCE 3.8. We are grateful to the Secretaría de Ciencias, Humanidades, Tecnología e Innovación (SECIHTI) for the financial support provided through the postdoctoral scholarship with CVU number 218255.}

\end{document}